\documentclass[twocolumn]{article}

\usepackage{graphicx,epsfig} 

\begin{document}
ZU-TH 22/01
\vskip2mm
\begin{center}
{\bf Conf\'erence du Club d'Information Scientifique}
\\
{\it Ecole Nationale Sup\'erieure des T\'el\'ecommunications}\\
 Paris (France), le 11 Juin 2001
\end{center}
\vskip2mm
\hrule
\vskip8mm
\noindent
\begin{center}
{\huge {\bf L'Univers sombre:}} 
\vskip2mm
{\huge {\bf Les premi\`eres mol\'ecules}}
\end{center}
\vskip8mm
\hrule
\vskip8mm
\noindent
{\bf \large Denis Puy}\footnote{Email: puy@physik.unizh.ch}
\vskip2mm
\noindent
{\it 
Institute of Theoretical Physik, Z\"urich
\\
and Paul Scherrer Institut, Villigen (Switzerland)}
\vskip4mm
\noindent
{\bf 1- Introduction}
\\
Titre \'etrange s'il en est. L'Univers sombre ou l'\^age sombre de l'Univers 
relate ni une \`ere tragique, ni un ant\'ec\'edent t\'en\'ebreux, ni une 
phase funeste de l'histoire de l'Univers ou un \'episode de la saga des 
{\it Star Wars }. 
Il s'agit d'une p\'eriode peu connue et donc ... obscure d'un lointain 
pass\'e de l'Univers. 
\\
Martin Rees fut le premier \`a sugg\'erer 
ce vocable pour d\'esigner l'interm\`ede de l'histoire de l'Univers qui a suivi 
le d\'ecouplage de la mati\`ere et du rayonnement, et pr\'ec\'ed\'e 
la formation des premiers objets. Durant cette p\'eriode, les conditions physiques 
de l'Univers permettent de penser que les premi\`eres mol\'ecules apparurent. Cette 
\'etape a \'et\'e peu \'etudi\'ee jusqu'\`a encore une d\'ecennie.
\\
La cosmologie s'appuie aujourd'hui essentiellement sur trois faits 
observationnels:
\begin{itemize}
\item L'expansion de l'Univers, qui \'eloigne les galaxies les unes des autres.
\item La composition de l'Univers en \'el\'ements l\'egers (environ 75 \% d'atomes 
d'hydrog\`ene, 25 \% d'atomes d'helium).
\item L'existence d'un fond de rayonnement cosmologique de caract\`ere thermique 
\`a environ 2.73 Kelvins.
\end{itemize}
A partir de ces observations, un mod\`ele d'\'evolution de l'Univers a pu \^etre 
\'etabli. C'est le mod\`ele du {\it Big Bang}, qui fait aujourd'hui l'objet d'un 
assez large consensus. Selon ce mod\`ele, l'Univers a connu une origine 
extr\^emement dense et chaude, suivie d'une expansion et d'un refroidissement. 
C'est au cours de ce refroidissement que la mati\`ere a, peu \`a peu, pris la 
forme que nous lui connaissons aujourd'hui. 
\\
Au cours de cet expos\'e j'analyserai, tout d'abord, les ph\'enom\`enes qui 
ont pr\'evalu avant la formation des premi\`eres mol\'ecules de l'Univers. 
Ainsi, je m'attacherai \`a une br\`eve description de la phase de formation 
des noyaux, ou nucl\'eosynth\`ese primordiale, suivie de celle de formation des 
premiers atomes ou recombinaison cosmologique. Cette derni\`ere aura des 
cons\'equences particuli\`erement importantes sur l'\'evolution de l'Univers, 
car c'est \`a cette m\^eme \'epoque que le rayonnement se d\'ecouple de la 
mati\`ere, conduisant \`a lib\'erer les photons dans un fond diffus. C'est 
pr\'ecis\'ement \`a cette p\'eriode que les premi\`eres mol\'ecules vont 
apparaitre dans l'Histoire de l'Univers. Nous tenterons alors d'analyser les 
cons\'equences de l'existence de celles-ci sur la phase ult\'erieure 
de formation des premi\`eres structures gravitationnelles.
\vskip2mm
\noindent
{\bf 2- La nucl\'eosynth\`ese primordiale}
\\
Le mod\`ele standard du {\it Big Bang} est bas\'e sur l'extrapolation de notre 
connaissance actuelle de l'Univers -expansion, existence du fond microonde- et 
sur celle de la physique des particules. Une description temporelle de l'Univers 
primordial, suivant son expansion, permet de d\'ecrire qualitativement les grandes 
\'etapes de son histoire \`a travers diff\'erentes transitions de phases successives:
\begin{itemize}
\item Une premi\`ere phase d'expansion extr\^emement brutale, 
appel\'ee {\it inflation}.
\item La transition de la {\it th\'eorie de grande unification}, quand l'Univers 
avait un 
\^age d'environ 10$^{-36}$ s et une temp\'erature de 10$^{28}$ K, qui d\'ecouple 
la force \'electrofaible et l'int\'eraction forte.
\item La transition {\it \'electrofaible} lib\'erant la force \'electromagn\'etique et 
l'int\'eraction faible, quand il avait un \^age d'environ 10$^{-10}$ 
s et une temp\'erature d'environ 10$^{15}$ K.
\item La transition {\it quark-hadron}, ou confinement des quarks en hadrons, 
quand il avait un \^age d'environ 10$^{-6}$ s et une temp\'erature d'environ 
10$^{13}$ K.
\end{itemize}
Apr\`es cette transition quark-hadron, l'Univers est constitu\'e essentiellement de neutrinos, d'anti-neutrinos, de positrons, d'\'electrons, de photons, de protons et de neutrons en \'equilibre 
thermodynamique (la soupe originelle). 
\\
A ces temp\'eratures, les neutrinos jouent un r\^ole stabilisateur dans cette agitation thermique. En effet 
absorb\'es et re\'emis sans cesse par les nucl\'eons, ces neutrinos 
transforment continuellement les protons en neutrons et inversement. Ces 
r\'eactions, gouvern\'ees par l'int\'eraction faible, maintiennent en \'equilibre 
une population de neutrons tout \`a fait comparable \`a celle des 
protons. L'Univers est alors opaque aux neutrinos. 
\\
Avec la d\'ecroissance de la temp\'erature l'\'energie des particules 
diminue progressivement, rapidement les neutrinos ne vont plus \^etre en 
mesure d'interagir avec les nucl\'eons. En dessous de la temp\'erature de 
$10^{10}$ K, l'Univers devient transparent aux neutrinos. Ce passage \`a 
la transparence va avoir des incidences particuli\`erement 
importantes sur la population des neutrons. A ce stade l'\'equilibre entre 
les protons et neutrons est rompu. Certes le neutron peut continuer \`a se 
d\'esint\'egrer en donnant naissance \`a un proton, un \'electron et un 
neutrino, mais en des temps de plus en plus longs. En revanche ce 
m\^eme neutron va r\'eagir beaucoup plus vite avec un proton et former un 
noyau de deut\'erium. 
C'est le d\'ebut de la phase de nucl\'eosynth\`ese primordiale. 
\\
Cette {\it brisure} de l'\'equilibre neutron-proton conduisant \`a la 
formation des 
premiers noyaux atomiques va s'intensifier \`a la temp\'erature de 10$^9$ K. 
Peu \`a peu d'autres noyaux tels que l'helium 3, l'helium 4 vont apparaitre 
puis interagir entre eux pour constituer un r\'eseau complexe de
r\'eactions nucl\'eaires caract\'erisant  le mod\`ele standard de la 
nucl\'eosynth\`ese. 
\\
Le deut\'erium apparait donc comme la premi\`ere {\it brique} de la synth\`ese des 
\'el\'ements l\'egers. Une fois pass\'ee la {\it chicane} du deut\'erium, les noyaux 
plus lourds peuvent se constituer: le deut\'erium est un passage oblig\'e dans la chaine 
des r\'eactions qui conduisent \`a la synth\`ese de l'h\'elium 4. Peu \`a peu 
l'\'energie du milieu diminue (par l'expansion), la barri\`ere Coulombienne 
entre les noyaux va \^etre de plus en plus efficace. La nucl\'eosynth\`ese 
va stopper au noyau de beryllium $^9Be$, une description plus compl\`ete 
de la nucl\'eosynth\`ese primordiale peut \^etre trouv\'ee dans Sarkar 1996 
ou Puy \& Signore 2001. 
\\
Un calcul pr\'ecis de quelques 150 r\'eactions nucl\'eaires connues 
conduit aux abondances primordiales suivantes (en unit\'e de densit\'e 
totale):
\begin{eqnarray}
& \bullet & H \sim 0.76 \nonumber \\
& \bullet & ^4He \sim 0.24 \nonumber \\
& \bullet & D \sim 4.3 \times 10^{-5} \nonumber \\ 
& \bullet & ^3He \sim 10^{-5} \nonumber  \\
& \bullet & ^7Li \sim 2.2 \times 10^{-10} \nonumber \\
& \bullet & ^9Be \sim \, {\rm trace}
\nonumber
\end{eqnarray}

Les tr\`es r\'ecentes mesures du fond de rayonnement cosmologique effectu\'ee 
par l'\'equipe conduite par Paolo de Bernardis (2000), \`a l'aide du ballon 
BOOMERANG lanc\'e en 1998 en Antarctique, semblent confirm\'ees ce mod\`ele 
standard de la nucl\'eosynth\`ese primordiale (Burles-Nollett-Turner 2001). 
\\
D'autres alternatives th\'eoriques furent tent\'ees. 
L'une concerne une nucl\'eosynth\`ese 
h\'et\'erog\`ene, o\`u lors du confinement quarks-hadrons certaines fluctuations 
de densit\'e pourraient donner lieu \`a former des bulles riches en neutrons 
ou en protons, et faciliter ainsi la production d'\'el\'ements lourds (voir Jedamzik \& Rehm 
2001). L'autre tentative a \'et\'e conduite par quelques th\'eoriciens de physique des 
particules qui ont sugg\'erer l'existence d'un quatri\`eme type de neutrino: 
le neutrino st\'erile, caract\'eris\'e par le fait que sa force d'int\'eraction est 
beaucoup moins importante que la classique int\'eraction faible. Une 
premi\`ere cons\'equence serait alors de modifier le rapport neutron-proton; 
param\`etre important pour la nucl\'eosynth\`ese primordial. Ce nouveau champ 
de recherche particuli\`erement actif au CERN de Gen\`eve pourrait conduire \`a 
de nouvelles valeurs d'abondance de noyaux primordiaux (voir Kirilova \& 
Chizhov 2001). Dans le cadre 
de ces mod\`eles, les noyaux plus lourds que les noyaux de nombre de masse 
sup\'erieur \`a 11 pourraient \^etre synth\'etis\'es. L'abondance 
primordiale des noyaux du $^7Li$ au $^{11}B$ serait \'egalement sup\'erieure 
\`a celle du mod\`ele standard. 
\vskip2mm
\noindent
{\bf 3- Chimie primordiale}
\\
Apr\`es cette courte p\'eriode de nucl\'eosynth\`ese, l'Univers est encore 
domin\'e par le rayonnement qui est compl\'etement coupl\'e \`a la mati\`ere 
par diffusion \'elastique. Cette diffusion ou diffusion Thomson des photons par 
les \'electrons se prolonge jusqu'\`a ce que l'Univers ait une temp\'erature 
comprise entre 5 000 K et 7 000 K. A ces temp\'eratures, la photoionisation 
des atomes devient peu \`a peu n\'egligeable. Les noyaux se {\it recombinent} 
alors avec les \'electrons libres pour former les atomes, c'est 
l'\'epoque de la recombinaison. 
\\
$He^{++}$, $He^+$ et $H^+$ se recombinent en suivant leurs potentiels 
d'ionisation d\'ecroissants. La mati\`ere change rapidement d'\'etat, 
passant d'un plasma ionis\'e \`a un gaz neutre. Quand le nombre 
d'\'electrons libres a suffisamment d\'ecru, la diffusion Thomson 
devient n\'egligeable: c'est l'\'epoque de la derni\`ere diffusion conduisant 
au d\'ecouplage d\'efinitif entre la mati\`ere et le rayonnement. 
Des mol\'ecules sont alors capables de survivre.
\\
Puisque le mod\`ele classique de nucl\'eosynth\`ese primordiale favorise la 
production d'un nombre limit\'e d'\'el\'ements l\'egers, la chimie apr\`es 
la recombinaison est essentiellement une chimie gazeuse de l'hydrog\`ene, du 
deut\'erium, de l'h\'elium et , \`a un degr\'e moindre, du lithium. 
Bien que, chronologiquement, les premi\`eres mol\'ecules form\'ees soient 
les ions mol\'eculaires $He_2^+$ et $HeH^+$, et bien qu'on ne connaisse pas 
toutes les sections efficaces de toutes les r\'eactions de cr\'eation et de 
destruction de mol\'ecules, les mol\'ecules primordiales les plus abondantes 
sont l'hydrog\`ene mol\'eculaire $H_2$ et les hydrures $HD$ et $LiH$ 
(voir Puy et al. 1993, Puy \& Signore 1999, Signore \& Puy 1999).
\\
La formation de $H_2$ commence avec le processus radiatif:
$$
H^+ + H \, \to \, H_2 + h\nu
$$
suivie d'un transfert de charge:
$$
H_2^+ + H \, \to \, H_2 + H^+.
$$
Une autre voie pour la formation de $H_2$ commence par un attachement radiatif 
et la formation de l'ion $H^-$:
$$
H+ e^- \, \to \, H^- + h\nu
$$
suivie par un d\'etachement associatif avec H:
$$
H^- + H \, \to \, H_2 + e^-.
$$
Quant \`a l'hydrure $HD$, forme isotopique de $H_2$, un transfert de 
charge d\'etermine l'ionisation de $D$:
$$
D + H^+ \, \to \, D^+ + H
$$
et la production d'une abondance significative de $H_2$ conduit \`a la 
formation rapide de $HD$:
$$
D^+ + H_2 \, \to \, HD + H^+.
$$
Enfin l'hydrure de lithium $LiH$ outre par classique association radiative 
(r\'eaction tr\`es lente):
$$
Li + H \, \to \, LiH + h\nu
$$
cette mol\'ecule se forme principalement \`a partir de l'ion mol\'eculaire 
$LiH^+$ par \'echange de charge:
$$
LiH^+ + H \, \to \, LiH + H^+ .
$$
Le r\'eseau complet comporte environ 120 r\'eactions.
\\
L'\'evolution des abondances sera rapide apr\`es le d\'ecouplage rayonnement-
mati\`ere. Au d\'ecalage spectral $z=300$ l'essentiel des mol\'ecules 
primordiales et ions mol\'eculaires sont form\'es, les abondances des 
mol\'ecules n'\'evoluent plus. l'expansion prive la chimie de r\'eactions 
collisionnelles du fait de la d\'ecroissance de la densit\'e. L'abondance est 
{\it gel\'ee}. Les plus r\'ecents calculs concernant l'abondance 
{\it finale} des mol\'ecules et ions mol\'eculaires donnent 
(voir Galli \& Palla 1998 ou Stancil et al. 1998) donnent en unit\'e de 
densit\'e totale au d\'ecalage spectral $z = 5$:
\begin{eqnarray}
& \bullet & H_2 \sim 1.1 \times 10^{-6} \nonumber \\
& \bullet & HD \sim 1.2 \times 10^{-9} \nonumber \\
& \bullet & LiH \sim 7.1 \times 10^{-20}  \nonumber \\
& \bullet & HeH^+ \sim 6.2 \times 10^{-13} \nonumber \\
& \bullet & LiH^+ \sim 9.4 \times 10^{-18} \nonumber 
\end{eqnarray}
\noindent
{\bf -4 Nuages mol\'eculaires primordiaux}
\\
Tous les sc\'enarii de formation de structures supposent la pr\'esence 
initiale de fluctuations de densit\'e.
\\
Une \'etude d\'etaill\'ee de l'influence des mol\'ecules primordiales 
sur les diff\'erentes phases de la dynamique a \'et\'e 
entreprise ces derni\`eres ann\'ees (Puy \& Signore 1997, 
Abel et al. 2000 ou Fuller \& Couchman 2000). Une 
approche simple est de consid\'erer une fluctuation de densit\'e 
sph\'erique homog\`ene et isol\'ee que l'on appelera 
{\it nuage} (Puy \& Signore 1995). On peut 
montrer qu'il y a, dans un premier temps  expansion, puis dans un second 
temps effondrement du nuage qui se d\'ecouple alors de 
l'\'evolution cosmologique du milieu. La pr\'erogative d'attraction de la 
gravit\'e va se mettre en action au point de stopper la croissance de cette 
{\it surdensit\'e}, et d'initier la phase d'effondrement. Pour un nuage de 
masse 10$^9$ fois la masse du soleil (caract\'eristique de la masse d'une galaxie), 
l'effondrement se produit quand la temp\'erature du rayonnement atteint 
150 K, l'Univers a alors un \^age d'environ 5$\times 10^7$ ans. Les mol\'ecules 
peuvent rester pr\'esentes dans l'effondrement. Il est clair que celles-ci 
ne vont pas rester longtemps thermiquement inerte. On peut l\'egitimement 
estimer que la mati\`ere en effondrement va devenir plus chaude que le 
fond microonde environnant. Les processus d'excitations collisionnelles suivis de 
d\'e-excitations radiatives (induite et spontan\'ee) des mol\'ecules vont 
\^etre alors pr\'epond\'erants. Les mol\'ecules vont agir comme {\it agent de 
liaison} entre les photons du fond de rayonnement et la mati\`ere 
en effondrement, et conduire \`a refroidir cette derni\`ere puisque le fond 
de radiation est plus froid que l'effondrement.
\\
On montra alors (Puy \& Signore 1997) que dans ce cadre, la mol\'ecule $HD$ 
est le r\'efrig\'erant le plus efficace. Une premi\`ere cons\'equence 
dynamique de l'existence de cette fonction de refroidissement est 
la possiblit\'e de d\'evelopper une instabilit\'e thermique. La description de 
ce m\'ecanisme peut s'effectuer tr\`es simplement. Si un syst\`eme refroidit, 
il s'\'ecarte donc de son \'etat d'\'equilibre thermique. La fonction de 
refroidissement responsable de ce changement thermique peut, sous 
certaines conditions, 
engendrer un \'etat d'instabilit\'e permettant la croissance des fluctuations 
de densit\'e, toujours pr\'esente dans un syst\`eme r\'ealiste. N\'eanmoins 
cette instabilit\'e thermique favorise uniquement la croissance de surdensit\'es 
\`a petites \'echelles \`a l'inverse de l'instabilit\'e gravitationnelle qui ne 
privil\'egie pas d'\'echelles particuli\`eres. On montra alors que la 
mol\'ecule $HD$ peut provoquer une instabilit\'e thermique autour de 200 K 
dans l'effondrement. Une telle instabilit\'e thermique peut alors conduire 
\`a la formation de sous-unit\'es dans le nuage en effondrement, puis \`a 
le fragmenter en nuages de plus faibles tailles et masses.
\\
Ce sc\'enario conduit \`a penser que des structures du type stellaire 
pourraient \^etre form\'ees avant les galaxies. L'\'evolution de ces \'etoiles 
primordiales de grandes masses serait alors rapide puis conduirait, apr\`es 
leurs explosions en supernovas, \`a contaminer le milieu en \'el\'ements 
lourds. Chakrabarti (2000) montr\`erent que dans 
ce cadre l'effondrement d'un nuage mol\'eculaire carbon\'e peut conduire \`a 
la formation significative d'ad\'enine $H_5C_5N_5$, une des mol\'ecules 
constitutives de l'ADN. Cette mol\'ecule pourrait \^etre apparu t\^ot dans l'histoire 
de l'Univers. De l\`a \`a dire que la vie serait tr\`es primitive dans 
l'histoire de l'Univers, il y a un monde (de connaissance \`a franchir). A 
plus modeste \'echelle, une chimie plus compl\`ete est actuellement \`a l'\'etude.
\\
N\'eanmoins la connaissance de la masse des sous-structures produit par la 
fragmentation reste encore tr\`es impr\'ecise. Les prochaines 
missions spatiales tels que HERSCHEL et le T\'elescope spatial de 
nouvelle g\'en\'eration NGST pourront donner une meilleure vision de 
l'Univers profond, et offrir des indications quant au processus de fragmentation. 
L'observation future des mol\'ecules primordiales peut \^etre 
s\'erieusement envisag\'ee avec les missions actuelles 
telles que FUSE ou ODIN, ainsi qu'avec le projet ALMA au Chili, 
d'interf\'erom\`etre d'un syst\`eme de 64 antennes de 12 m chacune, 
de l'organisation des observatoires sub-australs (ESO) qui permettra 
\'egalement d'obtenir une meilleure compr\'ehension de la formation des 
premiers objets. L'avenir s'annonce passionnant.
\vskip2mm
\noindent
\begin{center}
{\it Il faut une infinie patience pour attendre toujours 
\\
ce qui n'arrive pas...
\\
(Pierre Dac)
}
\end{center}
\vskip3mm
\noindent
{\bf Remerciements:} L'auteur remercie Guy Mizrahi et le comit\'e directeur 
du CIS pour l'invitation \`a pr\'esenter cet expos\'e, l'ensemble des 
participants pour leur accueil chaleureux, ainsi que Patrick Koch et Monique 
Signore pour des remarques didactiques.
\vskip4mm
\noindent
{\bf Bibliographie}\\
{\footnotesize
Abel T. et al. 2000 ApJ 540, 39
\\
Burles S. et al. 2001 ApJ 552, L1
\\
Chakrabarti S. 2000 Astr. \& Astroph. 354, L6
\\ 
De Bernardis P. et al. 2000 Nature 404, 955
\\
Fuller T., Couchman H.2000 ApJ 544, 6
\\
Galli D., Palla F. 1998 Astr. \& Astroph.  335, 403
\\
Jedamzik K., Rehm J. 2001 {\texttt astro-ph/0101292}
\\
Kirilova D. Chizhov M. 2001 {\texttt hep-ph/0102114}
\\
Puy D. et al. 1993 Astr. \& Astroph.  267, 337
\\
Puy D. et al. 1995 Comptes Rendus Acad\'emie \\
des Sciences 320, IIb, 619
\\
Puy D., Signore 1997 New Astr. 2, 299
\\
Puy D., Signore M. 1999 New Astr. Rev. 43, 223
\\
Puy D. Signore M. 2001 {\texttt astro-ph/0101157}
\\
Sarkar S. 1996 Rep. Prog. Phys. 59, 1493
\\
Signore D., Puy D. 1999 New Astr. Rev. 43, 185
\\
Stancil P. et al. 1998 ApJ 509, 1}
\vskip5.7cm
\noindent
{\bf Site sur les futures missions et projets:}\\
{\small
ALMA (projet ESO):
{\texttt http://www.mma.nrao.edu}
\\
\\
FUSE (mission NASA en cours): \hfill \\
{\texttt http://fusewww.gsfc.nasa.gov/fuse}
\\
\\
HERSCHEL (projet mission ESA): \hfill \\
{\texttt http://astro.estec.esa.nl/SA-general/Projects/First/first.html}
\\
\\
ODIN (mission CNES-SNSB-CSA-TEKES en cours): \hfill \\
{\texttt http://www.snsb.se/Odin/Odin.html}
\\
\\
NGST (projet NASA): \hfill \\
{\texttt http://ngst.gsfc.nasa.gov}
}
\end{document}